\documentclass[aps,prb,english,twocolumn,showpacs,preprintnumbers,amsmath,amssymb,floatfix,superscriptaddress]{revtex4-1}

\usepackage[T1]{fontenc}
\usepackage[utf8]{inputenc}
\usepackage{graphicx}
\usepackage{color}
\usepackage{amssymb}
\usepackage{babel}
\usepackage[normalem]{ulem}



\begin{document}

\title{Fluid sensitive nanoscale switching with quantum levitation controlled by 
$\alpha$-Sn/$\beta$-Sn phase transition }

\author{Mathias Bostr{\"o}m}
\email{mathias.a.bostrom@ntnu.no}
\affiliation{Department of Energy and Process Engineering, Norwegian University of Science and Technology, NO-7491 Trondheim, Norway}
\affiliation{Centre for Materials Science and Nanotechnology, University of Oslo, P. O. Box 1048 Blindern, NO-0316 Oslo, Norway}

\author{Maofeng Dou}
\email{maofengdou@gmail.com}
\affiliation{Center for Green Research on Energy and Environmental Materials, National Institute for Materials Science, Tsukuba, Ibaraki  305-0044, Japan }

\author{Oleksandr I. Malyi}
\email{oleksandrmalyi@gmail.com}
\affiliation{Centre for Materials Science and Nanotechnology, University of Oslo, P. O. Box 1048 Blindern, NO-0316 Oslo, Norway}

\author{Prachi Parashar}
\email{prachi.parashar@ntnu.no}
\affiliation{Department of Energy and Process Engineering, Norwegian University of Science and Technology, NO-7491 Trondheim, Norway}

\author{Drew F. Parsons}
\email{d.parsons@murdoch.edu.au}
\affiliation{School of Engineering and Information Technology, Murdoch University, 90 South St, Murdoch, WA 6150, Australia}

\author{Iver Brevik}
\email{iver.h.brevik@ntnu.no}
\affiliation{Department of Energy and Process Engineering, Norwegian University of Science and Technology, NO-7491 Trondheim, Norway}

\author{Clas Persson}
\email{clas.persson@fys.uio.no}
\affiliation{Centre for Materials Science and Nanotechnology, University of Oslo, P. O. Box 1048 Blindern, NO-0316 Oslo, Norway}
\affiliation{Department of Physics, University of Oslo, P. O. Box 1048 Blindern, NO-0316 Oslo, Norway}

\begin{abstract}
  We analyse the Lifshitz pressure between silica and tin
  separated by a liquid mixture of bromobenzene and chlorobenzene.  We
  show that the phase transition from semimetallic $\alpha$-Sn to
  metallic $\beta$-Sn can switch Lifshitz forces 
  from repulsive to attractive. This effect is caused by the difference in
  dielectric functions of $\alpha$-Sn and $\beta$-Sn, giving  both attractive 
  and repulsive contributions to the total
  Lifshitz pressure  at different frequency regions  controlled
  by the composition of the intervening liquid mixture. In this
  way, one may be able to produce phase transition-controlled quantum
  levitation in liquid medium.
\end{abstract}

\date{\today}

\maketitle

\section{Introduction}

Nanoelectromechanics have by now developed into quite a mature subject, 
where one deals routinely with separations between bodies of the order of a 
few nanometers. For these structures, the Lifshitz forces due to quantum 
fluctuations \cite{Casi,Dzya} become accordingly important.
This force  typically causes attraction between
surfaces and thus contributes to stiction, leading to collapse of
devices when surfaces approach each other.\cite{Serry98,Bhushan} 
It has  been shown, however,  that the Lifshitz force may be
repulsive,\cite{Dzya} or even in an intricate way change sign as
separation
increases.\cite{Zwol2010,Lamo2009,Ninh,bosserPRA2012,Rich2,Rich3,AndSab,Haux,EstesoJCP} 
The development of direct measurements of Lifshitz forces has provided a major thrust in 
searching viable systems for device engineering.\cite{Mund,Milling,Lee,Feiler}

Controlled nanomechanical devices could be designed by tailoring the
magnitude of the intermolecular interactions between surfaces.
Several studies have investigated how this may be achieved through
optical excitations and temperature dependent phase change
materials.\cite{Benassi,Lambrecht,Esquivel,Torricelli,Sedigh,Sedigh2,Sedigh3,Bimonte1,
Torricelli2,Klimchitskaya,Chen,Castillo,Bimonte2,Bimonte3,Galkina}
For example, crystallization of amorphous Ag-In-Sb-Te film has been 
predicted to increase the Lifshitz force up to 20$\%$ between gold and 
the alloy surface.\cite{Torricelli}  
However, a phase-transition controlled sign reversal of the Lifshitz 
forces is a novel idea that has not to our knowledge been proposed yet.

\begin{figure}[h]
\resizebox{0.85\columnwidth}{!}{\includegraphics*{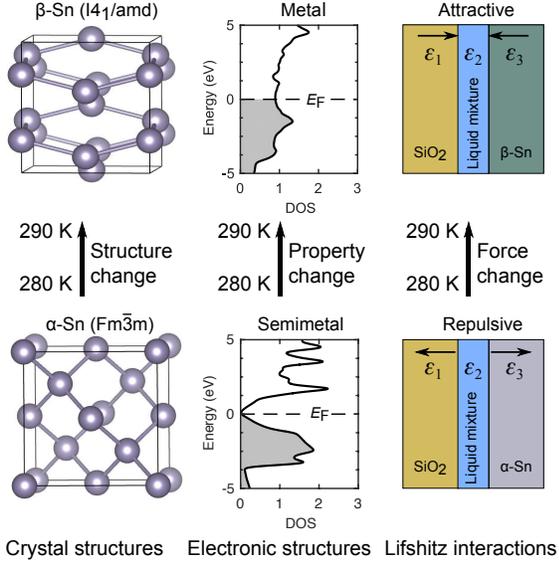}}
\caption{(Color online)  
Scheme of the three-layer system with switch from repulsive to attractive forces 
induced by the phase transition from $\alpha$-Sn to $\beta$-Sn. 
Left panel: the crystal structures of  the two phases. 
Middle panel: tin density-of-states (DOS) where $\alpha$-Sn is a semimetal 
with valence bands fully occupied, whereas $\beta$-Sn is a metal with
bands partially occupied across the Fermi level ($E_\text{F}$).
Right panel: the Lifshitz interactions between SiO$_2$ and tin surfaces 
across liquid mixtures at temperatures $T$ = 280 and 290\,K. 
$\varepsilon$ signifies the dielectric response in each layer.}
\label{Fig_tinStructure}
\end{figure}

In this paper, we introduce systems where a phase transition, induced by temperature
or other environmental factors, can switch the {\it sign} of the
Lifshitz force between surfaces of a phase transition material and a thin solid layer
across a very thin ($<$ 40{\,\AA}) liquid film.  
The model system we have in mind is shown in Fig.\,\ref{Fig_tinStructure}.
We propose to use a common phase transition material, solid tin, which has a phase 
transition temperature at $T$ = 286.5 K.\cite{Via,Iizumi} 
One of its two phases (grey tin; $\alpha$-Sn) is a semimetal, while  its other 
phase (white tin; $\beta$-Sn) 
is a metal; they have therefore very different dielectric responses. In order to obtain 
a phase-dependent transition from attractive to repulsive Lifshitz forces, 
the dielectric functions of the thin solid layer and the  intervening liquid must be 
close and must cross over.  One can achieve this requirement by constructing a system
with silica (SiO$_2$) as thin solid layer and a mixture of two (or more) 
liquids\cite{Lamo2009} whose dielectric function matches that of silica. 
A key element in our proposed model is the influence of the intervening liquid 
medium between the plates; the effect does not exist if the medium becomes replaced 
by a vacuum (or air). The transition distance from an attractive to 
a repulsive Lifshitz force occurs when the attractive and repulsive contributions to 
the total Lifshitz force from  different frequency regions exactly cancel.
Thus, the engineering requirements are apparent: the refractive indices of 
the two pure liquids can lie above and below the refractive index of one 
of the solid materials. It is then a matter of finding the right combination 
of the liquid mixture that yields the desired crossover of the dielectric 
functions of the liquid and the solid.
In this work, we choose a particular phase change material (i.\,e., tin) just for 
demonstrating the concept of switching, and we anticipate that this concept can be 
developed further utilizing also other types of metal/non-metal transition, for example 
by charge injection, chemical insertion, and magnetic phase transition.
These systems open up the possibility to make use of the Lifshitz effect as 
a switch, or actuator, that can be utilized in developments of 
microelectromechanical (MEMS) or nanoelectromechanical (NEMS)
systems,\cite{Serry98} as well as controlled low friction nanomechanical 
devices (Lifshitz repulsion leads to low friction between surfaces.\cite{Feiler})

\section{Theory}

Fundamental effects from the Lifshitz force is modelled for the three-layer system as
described in Fig.\,\ref{Fig_tinStructure}.
The retarded Lifshitz pressure $p(L)$, between silica and tin surfaces separated across a
liquid medium by distance $L$, is given as a sum over imaginary Matsubara 
frequencies ($\zeta_n=n 2\pi k_B T/\hbar$),\cite{Dzya}
\begin{equation}
p(L)=\sum_{n = 0}^{\infty}{'}\, (g^\text{TE}+g^\text{TM}),
\label{LifshitzPressure}
\end{equation}
where  prime on the summation sign indicates that the $n=0$ term shall be divided by two.
The  spectral functions for transverse electric and transverse magnetic modes $g^m$ 
($m =\text{TE}$ and $\text{TM}$) are
\begin{equation}
g^m =-\frac{k_B T}{2 \pi^2} \int d^2k \frac{   \gamma_2\,  r_{21}^m r_{23}^m  
e^{-2\gamma_2 L}}{  1- r_{21}^m r_{23}^m e^{-2\gamma_2 L}}.
\end{equation}
$r_{ij}^m$ are the reflection coefficients,
\begin{equation}
r_{ij}^\text{TM}=\frac{{{\varepsilon _j}{\gamma _i} - 
{\varepsilon _i}{\gamma _j}}}{{{\varepsilon _j}{\gamma _i} + {\varepsilon _i}{\gamma _j}}}
\quad \text{and} \quad
r_{ij}^\text{TE} = \frac{\gamma _i - \gamma _j}{\gamma _i + \gamma _j},
\end{equation}
where ${\gamma _i}(i \zeta_n) = \sqrt{k^2 +  {{\left( {\zeta_n
 /c} \right)}^2} {\varepsilon _i}}$.

To model the Lifshitz force accurately, a detailed knowledge of dielectric functions 
of all materials involved is essential.

\section{Modeling the dielectric responses of materials}

The primary materials considered in this work are tin ($\alpha$- and $\beta$-Sn) as the phase 
transition material, silica as the thin solid layer, and a liquid mixture based on bromobenzene (Bb) 
and chlorobenzene (Cb). In the next section, it will be demonstrated that by mixing liquid  
Bb with the less polarizable Cb, one obtains the necessary condition for a switch in the 
Lifshitz pressure from repulsion to attraction when $\alpha$-Sn undergoes a phase transition 
to $\beta$-Sn. 

\begin{figure}
\resizebox{0.85\columnwidth}{!}{\includegraphics*{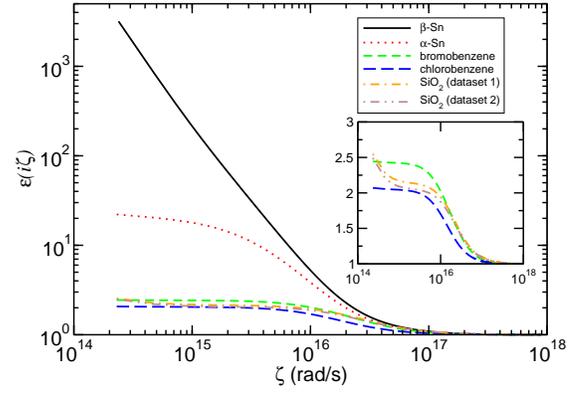}}
\caption{(Color online) 
The dielectric functions of bromobenzene (Bb), chlorobenzene (Cb), and silica (two slightly 
different datasets) are from van Zwol and Palasantzas (Ref.~\onlinecite{Zwol2010}).
Corresponding spectra for $\alpha$-Sn and $\beta$-Sn are obtained from DFT calculations. 
The average static dielectric constants are 
5.37, 5.75, 4.0, 27.2, $1.88\times10^5$,  
for Bb, Cb, SiO$_2$, $\alpha$-Sn, and $\beta$-Sn, respectively.
}
\label{Fig_dielectric}
\end{figure}

\subsection{Experimental dielectric functions of silica and liquid mixtures}

For the dielectric function of SiO$_2$ we consider two datasets (i.\,e., set 1 and set 2) 
given by van Zwol and Palasantzas.\cite{Zwol2010} 
The dielectric functions of Bb and Cb liquids are also taken from Ref.~\onlinecite{Zwol2010}.
The dielectric functions for the pure components  are shown  in Fig.\,\ref{Fig_dielectric}. 
The dielectric function of SiO$_2$ lies between those of Bb and Cb; see especially 
inset of Fig.~\ref{Fig_dielectric}. 

The mixing of two miscible liquids (Cb and Bb here)
adjusts the dielectric function of the intervening medium, whereby attractive and
repulsive contributions arising  from crossings of the dielectric functions of silica and
liquid will occur in different ways for $\alpha$-Sn and
$\beta$-Sn.(Cf. the remark of Lamoreaux\cite{Lamo2009} about the possibility to 'tune' 
the mixing such that the force becomes attractive at large separations and repulsive at 
short range.)
It is known that in a mixture of Bb and Cb the dielectric constant varies approximately 
linearly with the relative amount of each of the two components.\cite{LandBor} 
For the dielectric functions of  liquid mixtures, we use the Lorentz-Lorenz-like  
model\cite{Aspnes} with the susceptibility

\begin{equation}
\chi_2=\sum_{i=\text{Bb},\text{Cb}} V_{i} {\frac {\varepsilon_{2,i}-1} {\varepsilon_{2,i}+2}},
\label{LorenzLorentz}
\end{equation}
where $V_{i}$ is the volume fraction occupied by liquid $i$ component that has a dielectric function 
$\varepsilon_{2,i}$.
The dielectric function of the liquid mixture is then given by 
$\varepsilon_2=(1+2\chi_2)/(1-\chi_2)$.  

Since the calculated transition distances depend on how the dielectric functions are modeled,   
we have compared the model in Eq.~\ref{LorenzLorentz} with the volume average model that assumes 
a linear dependence of $\varepsilon_{2,i}$ on $V_i$. The two models describe rather similar 
dielectric spectra, and they both can give attraction to repulsion transitions. 
Inaccuracies of describing the exact dielectric responses can thus in an experimental setup 
be compensated by adjusting the liquid mixture to obtain the switching.   

\subsection{Calculated dielectric functions of tin}

For the two tin phases we modeled the dielectric functions within the density functional 
theory (DFT), employing the augmented plane wave method with local orbitals for Sn $d$-like 
orbitals (i.\,e., the APW+lo method) as provided by the WIEN2k package.\cite{DFT}  
The imaginary part of the dielectric tensor was calculated from the linear response 
of the momentum matrix elements describing the transition probability between 
occupied and unoccupied states.\cite{Draxl2006} 
Experimental lattice constants\cite{LandBor2} and two-atom primitive cells were used. 
The regular exchange-correlation potentials with the local density approximation (LDA) 
or the generalized gradient approximation (GGA) do not accurately describe tin, 
especially semimetal $\alpha$-Sn, due to overestimated hybridization between valence and 
conduction band states.\cite{Persson2006} Instead, 
we utilize the modified Becke-Johnson meta-GGA exchange potential combined with the 
LDA correlation potential. With a small k-mesh, we have verified a good density-of-states 
by comparing with a corresponding hybrid functional calculation. A dense k-mesh is however 
needed to describe details in the dielectric response accurately.\cite{Crovetto2016}
We, therefore, calculate it using the regular tetrahedron integration of the irreducible 
wedge of the Brillouin zone with $58\times58\times58$ k-mesh grids and an energy grid 
with step length of about 0.3\,meV. 
The plane-wave cutoff $K_{max}$ was determined from $K_{max}=8.4/R$  with near-touching 
the muffin-tin radii $R$. We have verified that the computed dielectric functions of both 
$\alpha$-Sn and $\beta$-Sn phases agree very well with ellipsometric spectra measured in 
the energy region 1.2 to 5.6 eV.\cite{Via} 
The corresponding dielectric functions $\varepsilon_3$ as functions of imaginary frequency  
were obtained from the Kramers-Kronig relation, where the intraband contribution for 
$\beta$-Sn assumed Drude broadening of 20\,meV. The dielectric functions of the two tin 
phases are displyed in Fig.\,\ref{Fig_dielectric}.

\section{Results: Fundamental effect}

\begin{figure}[t]
\resizebox{0.85\columnwidth}{!}{\includegraphics*{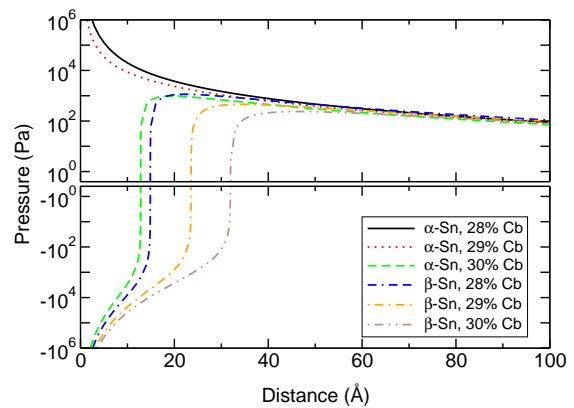}}
\caption{(Color online)  The Lifshitz pressure as a function of the
  distance $L$ between silica (dataset 1) and $\alpha$-Sn or $\beta$-Sn across a liquid 
  mixture (28$\%$, 29$\%$, and 30$\%$ chlorobenzene in bromobenzene), using the 
  dielectric functions from Fig.\,\ref{Fig_dielectric} and mixing according 
  to Eq.~\ref{LorenzLorentz}. 
  Positive values mean repulsion, negative values mean attraction.   }
\label{Fig_pressure_Cb}
\end{figure}

Both tin phases, interacting with silica across pure Bb, experience
repulsion at short separation distances. In contrast, across pure Cb,
an attractive short-range force is found between both phases of tin
and silica. One option to fine-tune the phase controlled quantum
levitation is to use a mixture of liquids tailored experimentally.\cite{Lamo2009} 
We show in Fig.\,\ref{Fig_pressure_Cb} the Lifshitz
pressure as a function of the distance between silica (dataset
1) and tin across three different liquid mixtures: Bb with either 
28$\%$, 29$\%$, and 30$\%$ Cb added. As expected, the interaction 
becomes attractive at longer distances as more of the less polarizable Cb 
is added to Bb. We observe that there is a strong phase transition
dependence in the sign of the Lifshitz pressure.  The range of
separation distances for the transition from attraction to repulsion
depends both on the tin phase and on the specific liquid mixture. 
It is possible to find the effect for thicker liquid films, but we focus
here on liquid mixtures that give transition from attraction to repulsion
in the limit of very thin liquid films.
Fig.\,\ref{Fig_pressure_Cb} suggests two alternative switching applications. 
On the one hand, switching in response to a phase transition 
(e.g.\ change in temperature),  and  on the other hand switching in response 
to a change in liquid composition (i.\,e., Cb content).

\begin{figure}
\resizebox{0.85\columnwidth}{!}{\includegraphics*{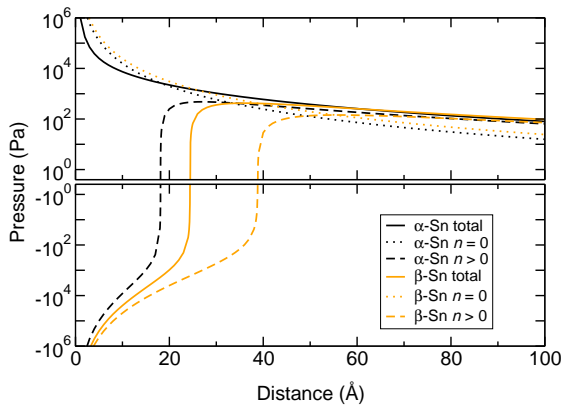}}
\caption{(Color online)  The total Lifshitz  pressure and its
  contribution from $n=0$ term and from $n>0$ terms in the Matsubara
  summation. The results are shown for $\alpha$-Sn and $\beta$-Sn as a
  function of their separation from silica (dataset 1) 
  across a liquid mixture (29.1$\%$ chlorobenzene in bromobenzene). }
\label{Fig_pressure_n}
\end{figure}

We illustrate the relative importance of the $n=0$ and $n>0$ contributions 
to the Lifshitz pressure in Fig.\,\ref{Fig_pressure_n}. We are facing a
situation where the zero-frequency term is as important as it is in systems 
involving water, but not for the reasons anticipated by standard wisdom 
(e.g.\,Ref.~\onlinecite{Pars1970} or Sect.~6.7 
in Ref.~\onlinecite{IsraelachviliBook2011}). 
In contrast to liquid water, it is
not a high dielectric constant or high dipole moment that drives the
dominance of the zero-frequency term in $\alpha$-Sn. Rather, the
dominance is due to a delicate cancellation between
negative and positive contributions in the $n>0$ terms.  We have reported
this kind of relationship between $n=0$ and $n>0$ previously in
ice-water systems.\cite{Bostrom2017} Cancellation of $n>0$ terms leads to repulsion, 
due to the  dominance of the zero frequency term, for both Sn phases for liquid films
thicker than 50{\,\AA}. For thinner liquid films the $n>0$ terms
dominate when tin is metallic, leading to an attraction. The $n=0$
term dominates when tin is semimetallic, leading to repulsion.  

\begin{figure}
\resizebox{0.85\columnwidth}{!}{\includegraphics*{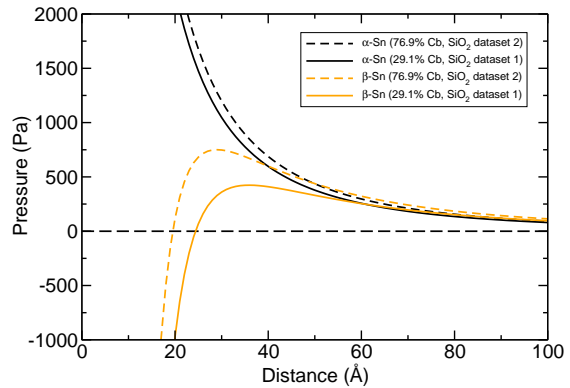}}
\caption{(Color online) The Lifshitz pressure as a function of the
  distance between silica dataset 1 and
  $\alpha$-Sn or $\beta$-Sn across a liquid mixture (29.1$\%$ chlorobenzene 
  in bromobenzene). For comparison, we also show the corresponding pressure using 
  an alternative dielectric function for silica dataset 2 
  combined with tin and a different liquid mixture(76.9$\%$ chlorobenzene 
  in bromobenzene).}
\label{Fig_pressure_SiO_diel}
\end{figure}

The liquids and mixing ratio need to be chosen and optimized for each system of 
phase transition considered. That is, with a certain mixture one can obtain repulsion 
for both phases, while another mixture yields only attraction. 
Between these two cases, one can find a range of mixing ratios suitable for a 
phase dependent nano-switch. 
To exemplify the sensitivity of the levitation with respect to changes
in the dielectric functions we present in Fig.\,\ref{Fig_pressure_SiO_diel}
the Lifshitz pressures for the two different silica materials 
(datasets 1 and 2). Each requires a different mixing ratio to work optimally as 
a phase-controlled nano-switch. 
The critical Cb concentration shifts from 29.1\% to 76.9\%.
However, the general behavior is similar after the critical Cb concentration 
has been tuned to optimise the attraction to repulsion distance. 
Many different silica materials (and similar materials, like mica or polystyrene)
will, when combined with a properly tuned liquid, provide further
examples where the phase transition from the semimetallic $\alpha$-Sn to
metallic $\beta$-phase changes the short-range Lifshitz interaction from
repulsion to attraction. When the same silica material is combined
with other liquid mixtures the sign of the interaction may be
independent of tin phase transition.  The reason, of course, is the
strict requirement to have a crossing of dielectric functions for the
specific silica material and the liquid.  When tin turns metallic, the 
interaction with the second surface turns more attractive (or less repulsive). 

\section{Results: Finite size silica layer}

While the previous section described the underlying physics of the interlayer interactions 
for the three-layer system tin/liquid/solid, this section discusses practical aspects in 
order to detect quantum levitation in liquid. We investigate  the thickness dependences 
on the solid layer using an extended thickness model.\cite{Bostrom2016} We consider therefore 
a vertically oriented layer-structure, and that the solid slab is able to move (or float) 
up and down in the liquid, and the slab feels the buoyancy pressure.  
This can thus be regarded as a four-layer system tin/liquid/solid/liquid containing a thin 
solid layer (typically SiO$_2$) with the finite thickness $d$ in a liquid 
(typically Bb in mixture with Cb). 
There is thus a thin film of liquid (thickness $L$) between tin and the solid, but also 
liquid above the solid slab. We will not allow the slab to float close to the liquid 
topmost surfaces, and therefore the liquid layer can be modeled with a semi-infinite 
thickness without any major loss in accuracy. Moreover, the bottom tin layer is still 
considered thick enough to be treated as semi-infinite.

\subsection{Thickness dependence of the Lifshitz pressure}

\begin{figure}
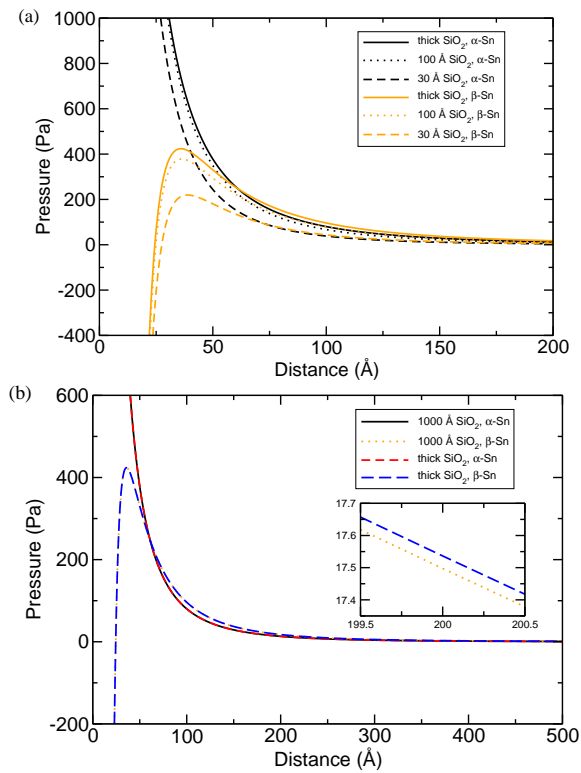

\resizebox{0.85\columnwidth}{!}{\includegraphics*{Figure6A.eps}}
\resizebox{0.88\columnwidth}{!}{\includegraphics*{Figure6B.eps}}
\caption{(Color online) (a) The Lifshitz pressure as functions of the separation 
distance $L$ between SiO$_2$ (dataset 1) and $\alpha$-Sn or $\beta$-Sn across a liquid 
mixture (29.1$\%$ chlorobenzene in bromobenzene) for the four-layer system with different 
thicknesses $d$ of the SiO$_2$ layer. Here, 'thick' implies the semi-infinite layer of 
SiO$_2$ used in the three-layer system (Fig.~\ref{Fig_tinStructure}). 
(b) Comparison of the Lifshitz pressures for 1000{\,\AA} thick SiO$_2$ layer and the 
semi-infinite layer of SiO$_2$.}
\label{Fig_pressure_thickness}
\end{figure}

We investigate the thickness dependence of the silica film in the Sn/liquid/SiO$_2$/liquid 
system containing a layer of SiO$_2$ with thickness $d$ in the liquid mixture 
29.1$\%$ chlorobenzene in bromobenzene. One can observe in Fig.~\ref{Fig_pressure_thickness}(a) 
that although the absolute values of the Lifshitz pressures depend on the thickness of 
SiO$_2$, the order of magnitudes of the pressure is comparable. The quantum levitation can 
be observed for all considered thicknesses. Moreover, when the thickness of the SiO$_2$
layer reaches 1000{\,\AA}, the distance dependence of the Lifshitz pressures overlaps in a 
large range of liquid layer thicknesses with that from the semi-infinite SiO$_2$ layer in the 
three-layer model in the main article, as shown in Fig.~\ref{Fig_pressure_thickness}(b). 
Thus, 1000{\,\AA} is large enough to be approximated as a macroscopic thickness.

\subsection{Buoyancy pressure}

\begin{figure}
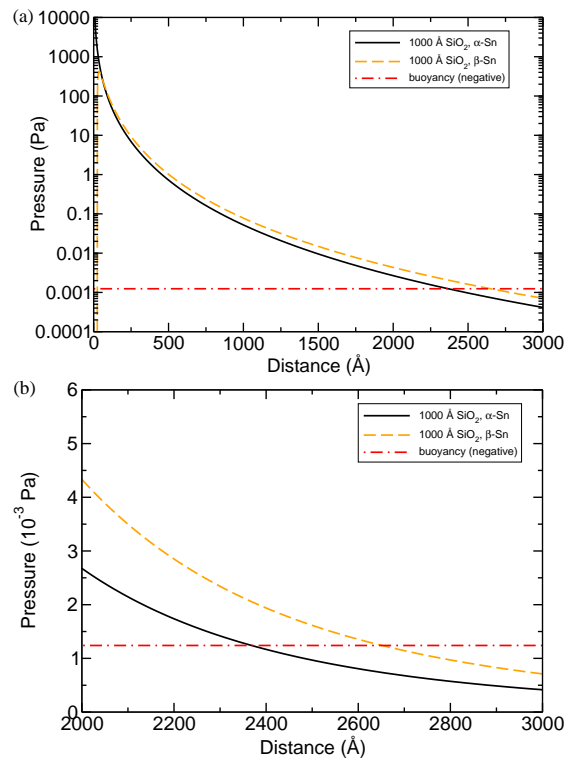

\resizebox{0.85\columnwidth}{!}{\includegraphics*{Figure7A.eps}} 
\resizebox{0.85\columnwidth}{!}{\includegraphics*{Figure7B.eps}} 
\caption{(Color online) (a) The Lifshitz pressure as a function of the separation 
distance between SiO$_2$ (dataset 1) and $\alpha$-Sn or $\beta$-Sn across a liquid 
mixture (29.1$\%$ chlorobenzene in bromobenzene) compared to the attractive buoyancy 
pressure (here, presented on a positive scale). 
(b) Magnification of the distance region where the repulsive Lifshitz and the attractive 
buoyancy pressures compensate each other.}
\label{Fig_pressure_buoyancy}
\end{figure}

The net buoyancy pressure $b$ on a SiO$_2$ slab in liquid due to gravity and difference in 
densities of the SiO$_2$ film and the surrounding liquid mixture can be estimated using 
$b = (\rho_{liquid}-\rho_{silica})\cdot gd$ where $g$ is the gravitational acceleration. 
With typical values for the densities of the liquid mixture, $\rho_{liquid}$, and 
of SiO$_2$, $\rho_{silica}$, and a thickness of the SiO$_2$ slab of $d=1000${\,\AA} 
the buoyancy pressure is $b \approx -1.2$\,mPa, where the negative sign indicates 
attraction. This value is negligible compared to the Lifshitz pressure at small separation 
distances ($L$ < 20{\,\AA}) where the quantum levitation occurs as shown in 
Fig.~\ref{Fig_pressure_buoyancy}(a). Figure~\ref{Fig_pressure_buoyancy}(b) demonstrates that 
the attractive buoyancy pressure can compensate the long-range repulsive Lifshitz 
contribution at large separations. Intriguingly, $\alpha$-Sn and $\beta$-Sn exhibit a 
noticeable difference in their respective equilibrium distances, where the net pressure 
due to the Lifshitz and buoyancy contributions vanishes; they differ by more than 200{\,\AA} 
which is obvious in Fig.~\ref{Fig_pressure_buoyancy}(b). However, although the effect 
is induced by the phase transition, it is not linked directly to the quantum levitation 
found for the small separation distances.

\subsection{Role of dielectric properties of interacting materials}

\begin{figure*}
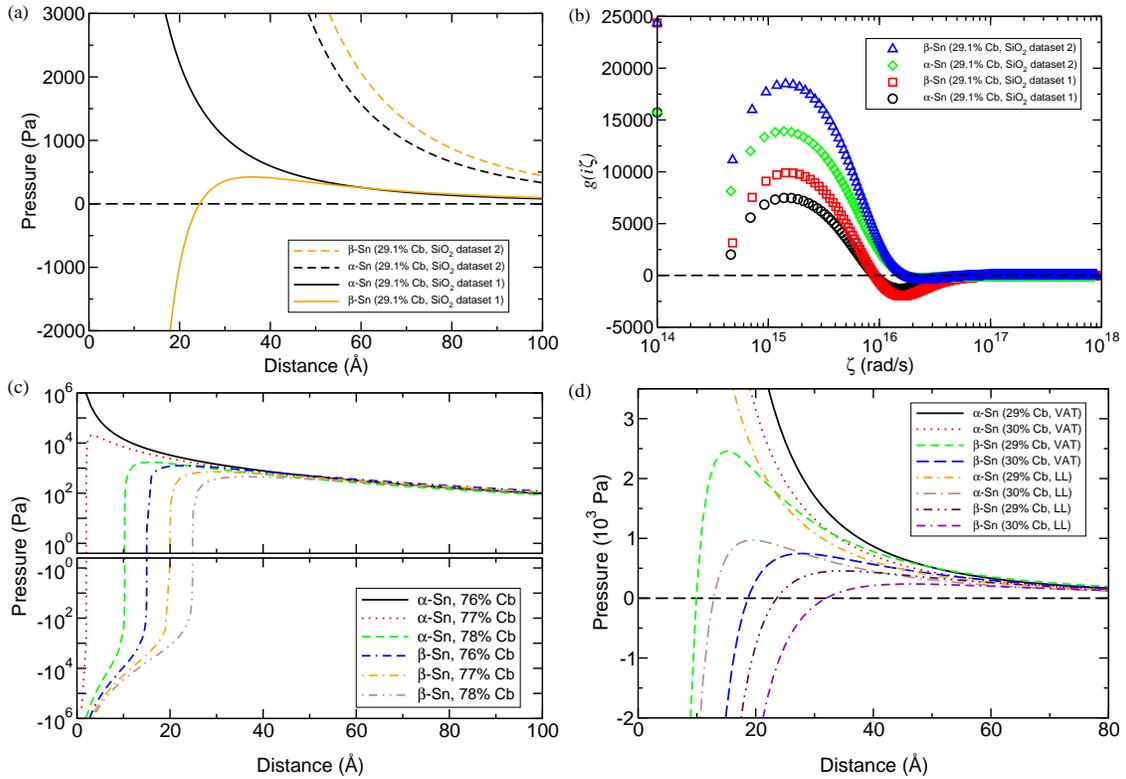

\resizebox{0.85\columnwidth}{!}{\includegraphics*{Figure8A.eps}}
\resizebox{0.85\columnwidth}{!}{\includegraphics*{Figure8B.eps}}
\resizebox{0.85\columnwidth}{!}{\includegraphics*{Figure8C.eps}}
\resizebox{0.85\columnwidth}{!}{\includegraphics*{Figure8D.eps}}
\caption{(Color online) (a) The Lifshitz pressure as a function of the separation distance 
between SiO$_2$ and $\alpha$-Sn or $\beta$-Sn across a liquid mixture of chlorobenzene and 
bromobenzene using two different parameterizations for the dielectric function of SiO$_2$, 
i.\,e., datasets 1 and 2. 
(b) Spectral functions revealing the contribution of each frequency mode to the Lifshitz 
pressures for the two different dielectric functions of SiO$_2$. The zero frequency term 
is divided by a factor of 2. 
(c) The Lifshitz pressure as a function of the separation distance between SiO$_2$ and 
$\alpha$-Sn or $\beta$-Sn across a liquid mixture of chlorobenzene and bromobenzene using 
SiO$_2$ dataset 2; this can be compared with Fig.~\ref{Fig_pressure_Cb}, where SiO$_2$ 
dataset 1 is used. 
(d) The Lifshitz pressure as a function of the separation distance between SiO$_2$ and 
$\alpha$-Sn or $\beta$-Sn across a liquid mixture of chlorobenzene and bromobenzene 
using SiO$_2$ dataset 1 with two different models to describe the dielectric function of the 
liquid mixture, namely the Lorentz-Lorenz-like (LL) and the volume average theory (VAT) models.}
\label{Fig_pressure_eps}
\end{figure*}

When the dielectric function for SiO$_2$ (dataset 1) is replaced with different 
parameterizations corresponding to a different SiO$_2$ sample (dataset 2), one can observe 
in Fig.~\ref{Fig_pressure_eps}(a) that the attraction to repulsion transition disappears. 
This effect is expected as noticeable variation in the dielectric properties has been reported 
in previous works.\cite{Zwol2010,Malyi2016} The difference between the two different SiO$_2$ 
samples (i.\,e., dataset 1 and dataset 2) may appear to be small (see Fig.~\ref{Fig_dielectric}) 
but Fig.~\ref{Fig_pressure_eps}(b) demonstrates that the spectral functions are very different. 
With the alternative dielectric function of SiO$_2$ (dataset 2), the repulsive contributions to 
the Lifshitz pressure are enhanced, and the attractive contributions to the Lifshitz pressure are 
reduced as compared to the corresponding results for the SiO$_2$ dataset 1. To obtain attraction 
for the interaction between the silica dataset 2 and tin, one must reduce the magnitude of the 
dielectric function of the liquid. This change can be achieved by increasing the ratio of 
chlorobenzene in bromobenzene, as described by Fig.~\ref{Fig_pressure_eps}(c). In the region 
between the different limits with the only repulsion and with only attraction, there is a ratio 
region where phase transition controlled attraction to repulsion transitions can occur. 
It is worth noticing that the utilization of different models for the dielectric function of 
liquid mixture can result in a variation of absolute values of the Lifshitz pressure. 
Nevertheless, the quantum levitation can still be achieved by making a small change in liquid 
ratio as demonstrated in Fig.~\ref{Fig_pressure_eps}(d).

\begin{figure}
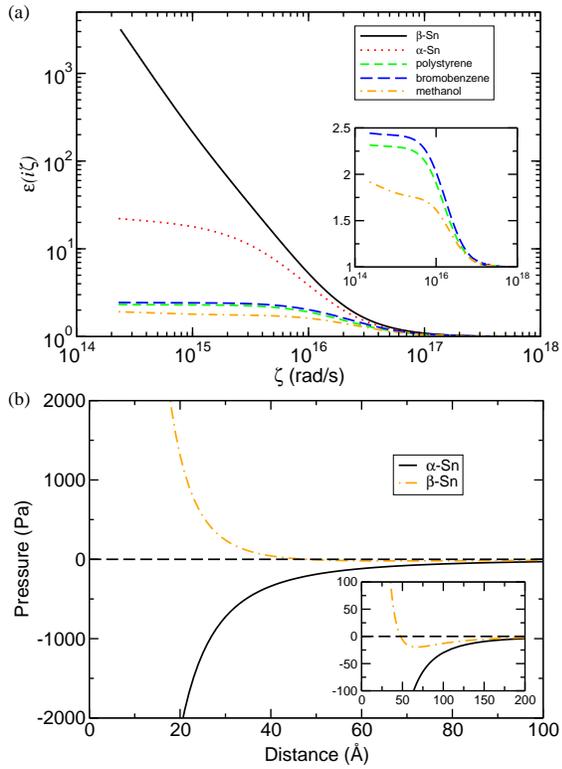

\resizebox{0.85\columnwidth}{!}{\includegraphics*{Figure9A.eps}} 
\resizebox{0.85\columnwidth}{!}{\includegraphics*{Figure9B.eps}} 
\caption{(Color online) (a) Dielectric functions of materials used in calculations of 
(b) the Lifshitz pressure as a function of the separation distance between polystyrene (dataset 1) 
and $\alpha$-Sn or $\beta$-Sn across a liquid mixture of 29$\%$ methanol in bromobenzene.
The static dielectric constant is 2.45 for polystyrene and 32.9 for methanol; 
data for the other materials are found in 
Fig.~\ref{Fig_dielectric}.
}
\label{Fig_pressure_polystyrene}
\end{figure}

Phase transition induced attraction to the repulsion of the Lifshitz pressure can also be 
observed for other systems of materials. In particular, it is found for the interaction of 
a 1000{\,\AA} thick polystyrene film with $\alpha$-Sn or $\beta$-Sn slab in a liquid mixture 
of methanol and bromobenzene; see Figs.~\ref{Fig_pressure_polystyrene}(a) and (b). 
The dielectric functions of polystyrene, methanol, and bromobenzene are also taken from 
van Zwol and Palasantzas's work.\cite{Zwol2010} 
The intervening liquid dielectric function needs to have 
a crossover with the dielectric function of one of the solid materials to obtain the levitation, 
but the liquid is a mixture of the two pure liquids whose dielectric functions can lie on 
either side of the solid [Fig.~\ref{Fig_pressure_polystyrene}(a)]. Thus, the engineering 
requirements are obvious: the refractive indices of the two pure liquids can lie above and 
below the refractive index of one of the solid materials. It is then a matter of finding 
the right combination of the liquid mixture that yields the desired crossover of the dielectric 
functions of the liquid mixture and the solid. Hence, it is possible to achieve quantum 
levitation controlled by the $\alpha$-Sn/$\beta$-Sn phase transition with combinations of 
different materials. Noticeable in Fig.~\ref{Fig_pressure_polystyrene}(b) is that here the 
$\beta$-Sn system yields more repulsion compared to $\alpha$-Sn. This reversal of behavior 
occurs due to the difference in frequency regions that give positive and negative 
contributions to the Lifshitz pressure. Since the zero frequency contributes repulsion, 
the resulting pressure for both systems is repulsive at distances beyond 760{\,\AA} for 
$\alpha$-Sn and 470{\,\AA} for $\beta$-Sn, where the lower frequency mode contributions dominate. 
Thus, there are two crossings of zero pressure, with multiple extreme points, in the pressure 
curves for the polystyrene-liquid-($\beta$-Sn) system.

\subsection{Thermal fluctuation}

\begin{figure}[h]
\includegraphics[width=7cm]{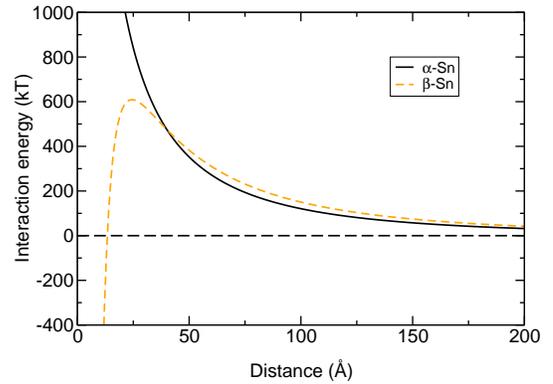}
\caption{(Color online) Interaction energy between SiO$_2$ (dataset 1) and 
$\alpha$-Sn or $\beta$-Sn across a liquid mixture (29.1$\%$ chlorobenzene in bromobenzene) 
as a function of the distance. The interaction energy is estimated from the pressure by 
conversion to a $0.1\times1\times1$\,$\mu$m slab of SiO$_2$. 
The temperatures are 280 and 290\,K for
$\alpha$-Sn and $\beta$-Sn, respectively.}
\label{Fig_interaction_energy}
\end{figure}

We identify two distinct ways in which the system is affected by thermal fluctuations. 
First, the contribution of the thermal fluctuations to the Lifshitz force is already 
accounted for\cite{Brevik2006} through the use of Matsubara frequencies $\zeta_n$ 
in Eq.~\ref{LifshitzPressure}. 
Second, the proposed system is also subject to classical thermal fluctuations arising 
from the kinetic energy of the surfaces at temperature $T$. Kinetic energy of the 
surface (a type of Brownian motion) causes the distance between surfaces to fluctuate. 
The stability of the system with respect to classical thermal fluctuations can be established 
by considering a finite contact area between tin and SiO$_2$ surfaces. The interaction energy 
is shown in Fig.~\ref{Fig_interaction_energy}, evaluated from the pressure assuming a 
1000{\,\AA} = 0.1\,$\mu$m thick SiO$_2$ slab with a 1\,$\mu$m$^2$ contact area.
The thermal kinetic energy is of the order of $k_BT$, thus as a rule of thumb, an interaction 
of more than 100\,$k_BT$ is robust with respect to thermal fluctuations, while an interaction 
weaker than 10\,$k_BT$ is reversible 
(not mechanically stable).\cite{Poortinga2001,Grasso2002,Chang2002} 
The repulsive barrier indicates that the attractive force is stable when the separation 
between $\beta$-Sn and SiO$_2$ is within $L \approx 20${\,\AA}. At the same length scale, the 
repulsive interaction found for $\alpha$-Sn exceeds 600\,$k_BT$, indicating stable repulsion. 

With these properties the system can be conceived as a trigger switch, initially set up for 
the $\beta$-Sn phase in close contact ($L < 20${\,\AA}) with the SiO$_2$ slab in the liquid
mixture. When a tin phase transition is triggered, $\alpha$-Sn repulsion pushes the SiO$_2$  
slab outwards. The device will need to be reset when tin is transformed back to the 
$\beta$-Sn phase, overcoming the 600\,$k_BT$ barrier either mechanically or by flushing with 
excess chlorobenzene for which the Lifshitz pressure is negative for all separation 
distances.
%%%%%%%%%

\section{Discussion}
 
It is of interest to apply a more comprehensive perspective and ask: will it be 
practically possible to apply this kind of phase transition system as an actuator?  

\begin{figure}
\includegraphics[width=7cm]{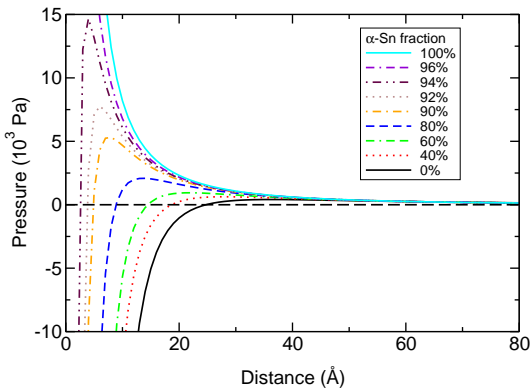}
\caption{(Color online) The Lifshitz pressure as a function of the
  distance $L$ between silica (dataset 1) and an $\alpha$-Sn/$\beta$-Sn mixture 
  across a liquid mixture (29.1$\%$ chlorobenzene in bromobenzene) for the 
  three-layer system (Fig.\,\ref{Fig_tinStructure}).
  The pressure is given for a partial tin phase transition at $T$ = 286.5 K 
  indicated by the fraction of $\alpha$-Sn coexisting with $\beta$-Sn. }
\label{Fig_pressure_partialTinTransition}%
\end{figure}

The timescale for the complete tin phase transition may be more than $10^4$ s.\cite{Nogita2013}  
In Fig.\,\ref{Fig_pressure_partialTinTransition} we show how the Lifshitz
pressure  varies during the process of conversion from $\alpha$- to $\beta$-Sn, 
when the two phases coexist.   
The dielectric functions of the $\alpha$-Sn/$\beta$-Sn mixture as well as of the 
liquid were evaluated using the Lorentz-Lorenz-like relation, Eq.~\ref{LorenzLorentz}, 
and we choose the three-layer system with semi-infinite SiO$_2$ thickness
as described in Fig.\,\ref{Fig_tinStructure}. 
The switch in pressure occurs after only 6\% conversion, at 94\% $\alpha$-Sn.  
It follows that the switch from repulsion to attraction can be achieved at a faster 
time scale than the time required for full conversion.

A second consideration for the practicality of this design of actuator
is frictional resistance from the fluid medium. An actuator executing
oscillations will necessarily be exposed to viscous drag forces. In
order to work properly, the decay time from the drag has to be much
shorter than the time needed for the phase transition. In this context we may
recall the instructive discussion given by Sedighi and Palasantzas:
\cite{Sedigh2} they replaced the upper plate by a metal sphere (gold)
with mass $M$ and radius $R$, elastically suspended in the
gravitational field, able to move vertically with velocity $\dot{z}$
under the combined influence of gravity, viscous drag from the
environment (in their case air), and Casimir force from the plate
beneath. The key difference from our case is that we are considering a
liquid-induced drag instead of an air-induced one. One can set up the
governing equation and from that estimate the viscous decay time. We
have gone through this calculation under our conditions, assuming a
gold sphere with $R=10~\mu$m, sphere velocity
$\dot{z}=3~\mathrm{mm}/\mathrm{s}$ typically, and viscosity as for
water. Under these conditions, the Reynolds number is very small,
hence the Stokes drag formula is applicable. We omit the details here, 
but the result is that the decay time becomes small, of the order of
milliseconds. This result is promising as the decay time is much
smaller than the time scale needed for the full tin phase
transition and the actuator can in our case be considered to be
reacting instantaneously. We point out that such devices are not
limited to the particular materials chosen, 
nor the type of metal/non-metal transition. Other combinations of  
materials and phase transitions can be chosen, tailored to the second 
surface in order to control switching and phase transition time. 
%---------------------------------------------------------

\vspace{5mm}

\section{Conclusions}
\label{conclusions}
As a conclusion, we have shown that by exploiting the combined effect
of (i) tin phase transition and (ii) fluid mixture composition, it
becomes possible to fabricate a switch operative at a moderate
temperature range.  A repulsive interaction is found in semimetallic
$\alpha$-Sn when the chlorobenzene  content is mixed with bromobenzene
(the specific critical concentration depending on the type
of silica). The repulsive interaction switches to attraction when
there is as little as 6\% conversion of $\alpha$-Sn to $\beta$-Sn. 
It also switches when the chlorobenzene content rises above the critical
concentration.   
We have verified that our model accurately can represent a 0.1\,$\mu$m 
thick SiO$_2$ slab with a 1\,$\mu$m$^2$ contact area, and that the 
buoyancy pressure and thermal effects then are negligible. 
For distances around $20${\,\AA} between Sn and SiO$_2$ surfaces, 
the repulsive Lifshitz free energy exceeds 600 $k_BT$ for $\alpha$-Sn, 
and its energy barrier is sufficiently large relative the thermal energy, 
indicating that the attractive interaction for $\beta$-Sn is stable. 
Hence, thermodynamically stable nano-switching can be achieved. 
We think that the idea is worth noticing for its possible
applications in nanomechanical systems and environmental sensors.

\begin{acknowledgments}
We acknowledge financial support from the Research Council of Norway (Projects 
221469 and 250346), and access to high-performance computing resources via 
SNIC and NOTUR.
\end{acknowledgments}

\end{document}